# Students Behavioural Analysis in an Online Learning Environment Using Data Mining

I. P. Ratnapala
Computing Centre

R. G. Ragel, S. Deegalla
Department of Computer Engineering

Faculty of Engineering, University of Peradeniya
Peradeniya, Sri Lanka

*Abstract* — The focus of this research was to use Educational Data Mining (EDM) techniques to conduct a quantitative analysis of students interaction with an e-learning system through instructor-led non-graded and graded courses. This exercise is useful for establishing a guideline for a series of online short courses for them. A group of 412 students' access behaviour in an e-learning system were analysed and they were grouped into clusters using K-Means clustering method according to their course access log records. The results explained that more than 40% from the student group are passive online learners in both graded and non-graded learning environments. The result showed that the difference in the learning environments could change the online access behaviour of a student group. Clustering divided the student population into five access groups based on their course access behaviour. Among these groups, the least access group (NG-41% and G-42%) and the highest access group (NG-9% and G-5%) could be identified very clearly due to their access variation from the rest of the groups.

## I. Introduction

The attitude towards learning, enthusiasm over new concepts and autonomous learning capability make an effective self-directed learner [2]. In a non-graded and non-obliged environment, students need a self determination to complete the course being a real self-directed learner.

The online Self Directed Learning (SDL) concept provides the opportunity to diversify the teaching of courses, cater more students with lesser staff, introduce multi-disciplinary courses and instigate inter-institutional resource platforms [1]. The major challenge of introducing on-line self-directed learning for university students is motivating the students' to complete the online courses successfully [3]. Since the learner participation is the crucial factor for the effectiveness of self-regulatory learning, a progress monitoring technique is required to observe the students behaviour with the system [2] .

A Learning Management System (LMS), which was developed to furnish online education, stores each record of the user interaction in the system as log entries. These log records makes a massive data storage, which contain all the information about the user interactions with the system [5].

The data mining technology can be applied to a large set of log records that are created from LMSs to understand the students' behaviour to improve the effectiveness of online education. This extended concept is known as Educational Data Mining (EDM). The educational data mining communities defined EDM as "an emerging discipline, concerned with developing methods for exploring the unique types of data that come from educational settings, and using those methods to better understand students, and the settings which they learn in"[1].

Self-Learning Online Short Courses (SLOSC) can be established in the educational institutes under the SDL concept to diversify the teaching courses, to introduce multi-disciplinary courses and to enhance the course curriculum with existing resources.

The aim of this research was to discover the students' online learning pattern for successful and sustainable installation of SLOSC in an educational institute as a substitute for traditional classrooms using a comparative analysis of the students' behaviour between non-graded and graded online learning environments.

## II. Related work

The SDL concept is being attracted by the researchers with the proliferation of online degrees all over the world.

Chou and Liu [19] presented a framework that addressed the relationship between the learner control and learning effectiveness in an online learning environment, which contains four categories: learning achievement, self-efficacy, satisfaction and learning climate. Therefore, the SDL concept should satisfy the learner requirement by providing an interesting environment to achieve the goals.

Song and Hill [18] proposed a framework to understand the SDL concept in an online environment. As per their findings, SDL needs to build the interaction between the learning environment and the learner. The learning environment includes the resources, course structure, collaboration and support and the learner is consisted with personal attributes and individual learning process. The on-line courses should pass three stages of design, support and outcomes for a sustainable SDL environment development and a learner should engage with the learning environment in a meaningful cognitive thinking to accomplish the desired goal with satisfaction.

Applying data mining techniques for the course log data of Course Management Systems (CMS) is an emerging area, with which the researches tend to carry out several empirical studies. Romero et al. [11] and Aher and Lobo [12] have defined that all the data mining techniques such as visualisation, clustering, classification, association rule mining, sequential pattern mining and text mining can be used to extract patterns through the CMS access data.

Barnard [13], Steffens et al. [14] and Sharma [15] have assessed self-regulatory on-line learning through questionnaires. They have indicated that self-learning is a skill, which students have to improve with proper guidance and the students tend to be more active self-learners at the on-line course environment with frequent assessments, quick instructor support, informative interactive content and direct student interactions.

---
[1] http://www.educationaldatamining.org/

Hung and Zhang [4] have done a similar study to reveal the undergraduates' online learning behavioural patterns and proposed a predictive model of user performance using the decision tree technique. They have indicated that the majority of the students were passive learners and only tend to access e-materials, but did not seek any peer collaborations. However, the few active learners showed a high performance level. They have proposed a decision tree for predicting performance.

As it has been mentioned in previous studies, SDL is an effective way of knowledge dissemination if the students successfully adopt it. The past studies revealed that the EDM concept can also be used to identify the learner behaviours in an online environment. The students groups are not properly analysed using EDM concept to reveal the guidelines for proper establishment of the SDL in on-line learning environments. This study focuses on quantitatively analysing the online learning behaviour of a student group in two different learning environments to reveal the similarities and differences of their learning behaviours. The revealed information based on the analysis will be considered for sustainable SDL establishments in educational institutes.

## III. BACKGROUND

### A. Moodle Database Structure

Modular Object-Oriented Dynamic Learning Environment (MOODLE), one of the popular[2] learning management systems that provide the most comprehensive e-learning tools [6] with a comprehensive database schema which contains around 230 tables. There are 50 core tables, which controls basic functions of the system [7].

Systems logs in Moodle are maintained by three tables "*log*", "*log_display*" and *log_queries*. The log data are stored in the "log" table and two other tables are used to display the data through the system reports [7].

The "log" table in a Moodle database records entries of each user interaction with the system. It connects with the user table, course table, course module table, resource and the activities tables such as quiz, forum, assignment, etc. to store all the log details of the system for each user interaction.

### B. Data Mining

Data mining or knowledge discovery in a database can be used to disclose the hidden relationship among the records. Clustering is one of the data mining techniques, which can be used to discover the new categories, which share the similar interest. In clustering the similar instances are grouped together. Different clustering algorithms can be used to separate the instances into given number of clusters. Intra-cluster homogeneity measures the validity of the cluster as the internal index [8].

### C. K-Means algorithm for clustering

Among the available clustering methods, K-Means algorithm is generally used to divide learners into natural groups based on their behaviour for a larger dataset. In the K-Means clustering method, the number of clusters, denoted by K is needed to be predefined to apply the technique [8].

This is one of the simplest and the most used unsupervised learning algorithm for clustering. The biggest problem in K-Means is the finalising of the optimum number of clusters. There are several simple and complex methods available to determine the K-Value. Sum of Squared Error (SSE) which defined "the error" as the distance to the nearest cluster for each point and within the cluster SSE is the most used and the simplest factor to evaluate the correctness of the clustering results. The number of clustering is inversely proportional until the SSE reaches the nadir. Therefore, the number of clusters can be determined considering the objective of the study and the SSE values [8].

WEKA [16] and R [17] were used for this study due its availability to download as open source software and compatibleness with CSV files.

### D. Educational Data Mining

EDM is capable of revealing system usage behaviours using data mining techniques. The clustering technique can be used to characterise the learner's behaviour and group them based on the behavioural similarities. Especially in e-learning, clustering is used as a learner diagnostic method to promote on line learning [11]. The process of clustering LMS log data should follow several steps such as; goal definition, data collection, data pre-processing, application of relevant DM technique and result interpretation [8].

## IV. RESEARCH METHODOLOGY

This research was conducted to perform an in-depth behavioural analysis of the students in the online e-learning environment. In this study, the data were obtained from two courses offered to 412 engineering students' group. The structures of the courses are as listed in Table I.

TABLE I. THE STRUCTURE OF THE COURSES

| Course 1 | Course 2 |
|---|---|
| Offered for fresh undergraduates | Offered at the 2$^{nd}$ semester |
| Introductory course, not part of the undergraduate curriculum | Core course under the undergraduate curriculum |
| Not graded | Graded |
| Teachers developed the syllabus | Department developed the syllabus |
| Six hours of lectures and lab classes | 45 hours of lectures and lab classes |
| Materials mainly target SDL | Materials to support for the lectures |
| Few instructors managed the forums and the lab classes | More instructors were assigned to manage the lab classes |
| Forum participation was not monitored and not given marks | Forum participation was monitored and marks are given |
| Text and video Self-learning resources were available | No self-learning materials were provided. |

The students' activities on the following Moodle modules in above courses (Table I) were considered for this study:

---

[2] http://www.capterra.com/learning-management-system-software/#infographic

a. *Forums View*: View log records are created when the students just see the forum topics or the discussion. The records related to "Forum view" or "discussion view" actions in the log table are being considered for this parameter.
b. *Forum Participation*: The students' actual (active) participation was considered under this parameter. The data have been considered when the action is equal to "add discussion" or "add post" or "update post" in the log table under the forum module.
c. *Resources (pages, files and URL)*: The resources are being added by the teacher for students to view. All the considered resources are non-interactive and contain only the information, references or course contents for the students to view.

The username had been created according to a pre-defined format. That username format reflects the studentship and the student group of each user. This criterion always provides comprehensive filtering options in the data refinement process in EDM.

The students were given a comprehensive training on using the e-learning system before the course commencement. Both courses were developed in a Moodle e-learning system and were delivered under the supervision of the instructors and the lecturers. Instructors were assigned to provide on-request support. The courses were conducted in scheduled durations.

The data processing of this study was conducted under several steps as depicts in Fig. 1.

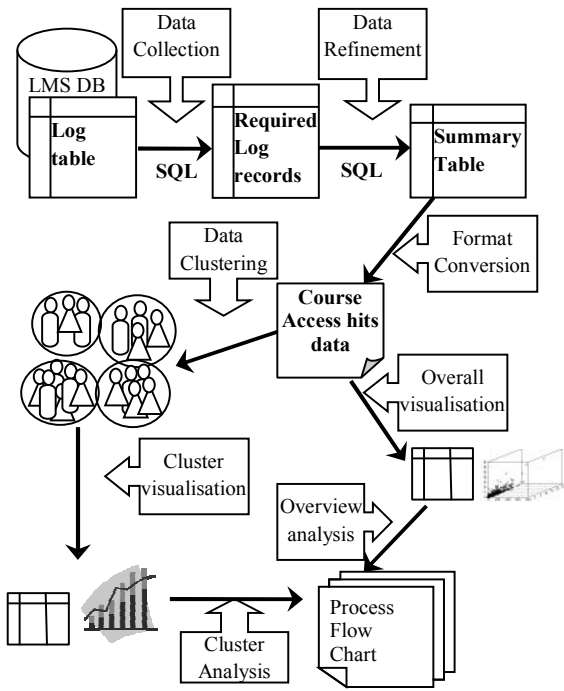

Fig. 1. The Process of Data collection, preparation and analysis

A. *Data collection, summarisation and refinement*

The log table of the Moodle system is queried to extract the specific course related data in the selected two courses separately. There were 31,735 log records for course 1 and 59,129 log records for course 2. The data were directly updated to two new tables, which contained all the users' details (username) as records which were used as access summarisation table. The two access summarisation tables were named as *tblsumdata_course1* and *tblsumdata_course2*. Both tables had the same table structure, which contains 5 data fields as shown in the Fig. 2. The primary key of the table is an auto increment number and the student no field was filled from the user table.

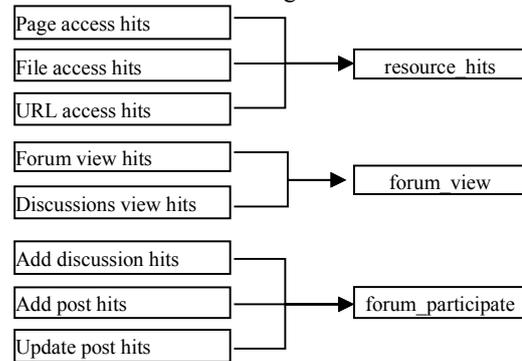

Fig. 2. Data summarisation table – Course1

Data have been added to the summarisation table using the UPDATE SQL queries selecting the records, which satisfy the conditions as shown in the Fig. 3.

```
Page access hits  ┐
File access hits  ┼──► resource_hits
URL access hits   ┘

Forum view hits       ┐
Discussions view hits ┴──► forum_view

Add discussion hits ┐
Add post hits       ┼──► forum_participate
Update post hits    ┘
```

Fig. 3. Data selection conditions

The data of completed summary tables of two courses have been exported into Comma Separated Value (CSV) file format to create two CSV data files.

B. *Converting to the WEKA supported format*

The two CSV files were converted to Attribute-Relation File Format (ARFF) through ARFF viewer in WEKA, which can load the dataset for applying clustering algorithm through WEKA.

C. *Data clustering*

The K-Means algorithm was applied to the two datasets using the Euclidean distance method. Here K is the numbers of clusters seek in the dataset. The algorithm has been applied for the dataset for different K value from 2 to 6 to pick the optimum K value.

D. *Visualisation*

The CSV files were directly used in R to summarise and visualize the overall data. The obtained clustering data from WEKA have been copied to the Microsoft Excel to generate the required graphs to visualise the clustered data.

E. *Analysis*

Overall data visualization and cluster data visualisation were used to perform a student behavioural analysis of both environments for the purpose of identifying process for successful installation of SLOCS in an educational institution.

## V. RESULTS AND DISCUSSION

### A. Overall Data

*1. Course 1 Data*

Table II depicts the overall result for course 1 and the mean values of each factor is considerably low (4.92, 7.77 and 0.90).

TABLE II. COURSE 1 ACCESS HITS SUMMARY

|  | Total | Mean | Median | SD |
|---|---|---|---|---|
| resource_hits | 2028 | 4.92 | 4 | 3.59 |
| forum_view | 3201 | 7.77 | 5 | 10.09 |
| forum_participate | 370 | 0.90 | 1 | 1.33 |

The graph (Fig. 4) shows that the majority of data points are much closer to the initial point and very few data points showed a significant elevation.

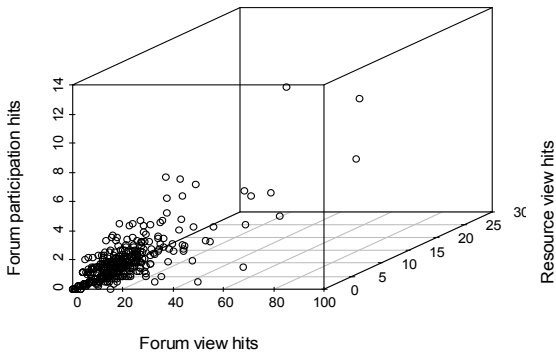

Fig. 4. 3-D graph of access hits for Course 1

*2. Course 2 Data*

Table III depicts that the students' forum view hits (M=38.68, SD=59.27) and forum participate hits (M=2.01, SD=3.2) in course 2 were significantly higher compared to the forums view hits (M=7.77, SD=10.09) and forum participate hits (M=0.9, SD=1.22) in the course 1 as in Table II.

TABLE III. COURSE 2 ACCESS HITS SUMMARY

|  | Total | Mean | Median | SD |
|---|---|---|---|---|
| resource_hits | 3335 | 8.09 | 6 | 6.98 |
| forum_view | 16635 | 40.38 | 28.5 | 46.68 |
| forum_participate | 2234 | 5.42 | 4 | 6.33 |

Fig. 5 shows a higher tendency in students' access in course 2 and the number of students who have obtained higher number of hits rate for considered factors have also being increased. The grading environment and the students experience with the e-learning system may have caused the higher hits rate of course 2. Personal attribute, process and the context are directly influenced for successful SDL. It is required to evaluate learners' requirement for an intrinsic motivation for SDL. Certification, grading, prior experience, language competence and pre-knowledge will motivate the students for SDL [18]. The learning context refers to the learning environment such as instructor support, quality of the materials, course design and the system user friendliness which can do a great impact on SDL.

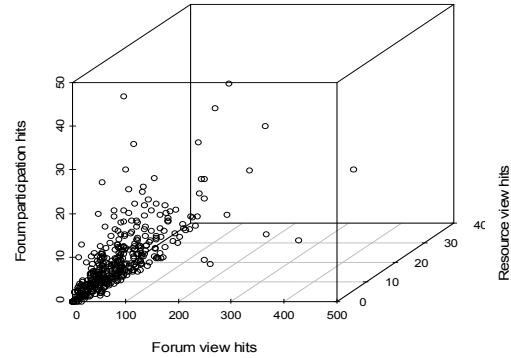

Fig. 5. 3-D graph of access hits for Course 2

### B. Data clustering

*Selecting the K-value for clustering*

The Sum of Squared Error (SSE) values for different K (number of clusters) values are shown in Table IV.

TABLE IV. SUM OF SQUARED ERROR FOR DIFFERENT K VALUES

|  |  | K-Value | | | | |
|---|---|---|---|---|---|---|
|  |  | 2 | 3 | 4 | 5 | 6 |
| SSE | Course 1 | 11.37 | 8.17 | 7.02 | 5.54 | 5.19 |
|  | Course 2 | 16.82 | 12.13 | 9.7 | 8.0 | 7.23 |

The SSE values when the K=2 is comparably high and it is getting lower when K is increased. When compare to other values, the minimum SSE deviation can be found between 5 and 6, which is less than 1. Therefore, the results obtained for when the K=5 have been considered for the evaluation.

*1. Course 1 Results*

K = 5 and SSE = 5.54

TABLE V. INSTANCE DISTRIBUTION OF CLUSTERS-DATASET 1

|  | Total | Cluster Numbers | | | | |
|---|---|---|---|---|---|---|
|  |  | 0 | 1 | 2 | 3 | 4 |
| No of Instances | 412 | 139 | 67 | 9 | 29 | 168 |
| Percentage | 100% | 34% | 16% | 2% | 7% | 41% |

TABLE VI. MEAN VALUES OF INPUT PARAMETERS-DATASET 1

| Input Parameters | Total hits | Mean Value | | | | | |
|---|---|---|---|---|---|---|---|
|  |  | Total | Cluster No | | | | |
|  |  |  | 0 | 1 | 2 | 3 | 4 |
| Resource hits | 2028 | 4.92 | 6.70 | 3.88 | 8.33 | 12.72 | 2.34 |
| Forum view hits | 3201 | 7.77 | 4.73 | 15.94 | 55.56 | 13.13 | 3.54 |
| Forum participate hits | 370 | 0.90 | 0.41 | 2.13 | 6.33 | 1.17 | 0.47 |

The course 1 result shows that only 9% (cluster 2 & 3) from the group have accessed all the modules above the total mean value while 41% (cluster 4) have accessed all lower to the mean value. Therefore, cluster 2 and 3 are consisted of the most active self-learners while cluster 4 becomes passive online learner group in non-graded environment. 34% fairly accessed the resources but showed a very poor participation on discussions. These students must develop communication skills and language competence. The cluster 1 students' contribution for the discussions is fairly accepted but the resource access tendency is not adequate. The lack of interest for the subject matters, the resources quality and the complexity might be some reasons for lowering their access hits for resources [19].

*2. Course 2 Results*
K = 5 and SSE = 5.04

TABLE VII. INSTANCE DISTRIBUTION OF CLUSTERS-DATASET 2

|  | Total | Cluster Numbers | | | | |
|---|---|---|---|---|---|---|
|  |  | 0 | 1 | 2 | 3 | 4 |
| No of Instances | 412 | 79 | 19 | 103 | 38 | 173 |
| Percentage | 100% | 19% | 5% | 25% | 9% | 42% |

TABLE VIII. MEAN VALUES OF INPUT PARAMETERS-DATASET 2

| Input Parameters | Total hits | Mean Value | | | | | |
|---|---|---|---|---|---|---|---|
|  |  | Total | Cluster No | | | | |
|  |  |  | 0 | 1 | 2 | 3 | 4 |
| Resource hits | 3335 | 8.09 | 5.28 | 13.53 | 12.11 | 23.32 | 3.05 |
| Forum view hits | 16635 | 40.38 | 67.18 | 159.48 | 34.85 | 48.21 | 16.62 |
| Forum participate hits | 2234 | 5.42 | 10.58 | 24.21 | 3.79 | 4.92 | 2.08 |

In course 2, there are only 5% (cluster 1) who has got higher mean values for all the three input parameters and 42% got lower values for all three (cluster 4). In cluster 0 (19%), students have used the forum for obtaining marks not with the cognition since they got a lower mean value for resource hits. The students who are grouped as cluster 2 and 3 (34%) with higher values in access resources and viewing forum and low values in forum participation depicts that they are lacking in communication skill and having problem with the written language. The highest student tendency to forum view (16635) depicts the highest teacher participation for the forums since the student forum participation (2234) is comparatively low compare to their forum view.

The above results of the two courses have been graphed for the comparison in Fig. 6 and Fig. 7.

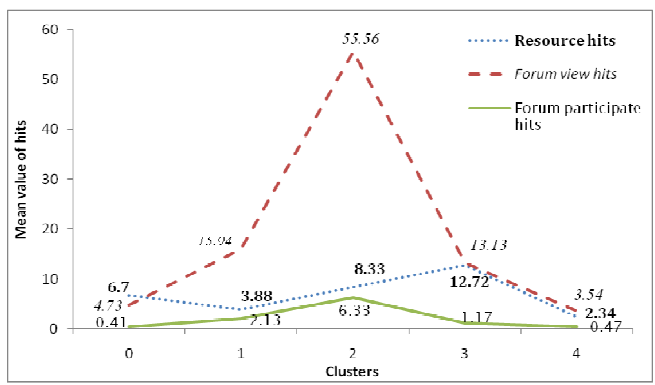

Fig. 6. Mean value distribution of each cluster in the Course 1

As shown in the Fig. 6 and Fig. 7, there is no similar mean distribution among clusters in the selected courses. The mean values of forum view and forum participation parameters of course 2 are significantly higher than the relevant figures of course 1.

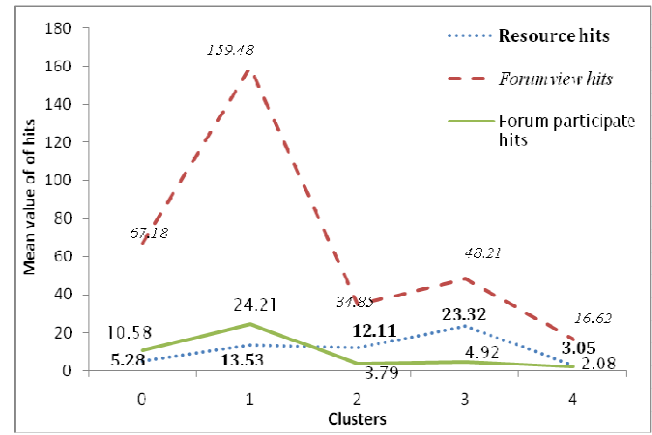

Fig. 7. Mean value distribution of each cluster in the Course 2

Resource access behaviour has reached the second popular parameter in course 1, while it is having an inverse relationship with forum participation in course 2. Graphs in Figures 6 and 7 depict that the grading and obligatory environment have motivated the students to access course 2 better. Even though, the forum participation was given a mark for students; 3 clusters (2, 3 and 4) are having the lowest value for forum participation.

The students' distribution among the identified clusters in both courses is graphed in the Fig. 8.

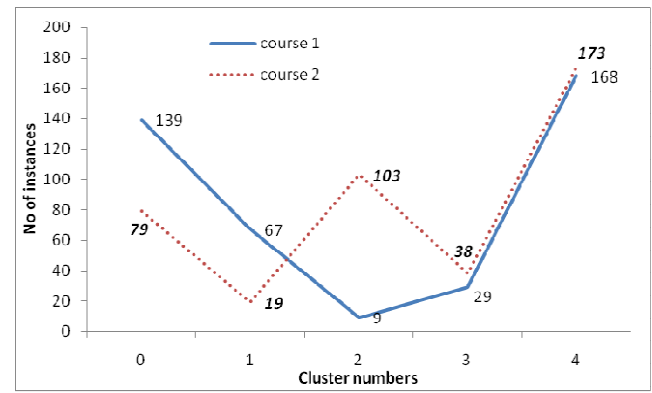

Fig. 8. Distribution percentage among groups in the courses

The Fig. 8 depicts the highest number of instances are in cluster 4 in both courses and it represent the passive online learners in the group. The difference in the learning environment has only increased the passive learning group by 5 numbers (course 1- 168 and course 2-173). However, the most active user percentage which refers the group who had higher mean value for all the parameters, has been decreased in course 2 (cluster 1) by 4% than course1 (cluster 2 and 3) since the students mostly focused on forum participation than accessing the learning materials in the course 2 environment.

Based on the above data analysis, the following information could be revealed about the selected student group.

- Nearly 40% (cluster 4 in course 1 and course 2) of the students are passive online learners those who do not perceive the usefulness of self-learning.

- Less than 10% (cluster 2, 3 in course 1 and cluster 1 in course 2) students can be considered as active learners in the students' group and they are the only real knowledge seekers in the group.

- Entire student group prefers to obtain grades and certification rather than acquire new knowledge (6 times mean value increment in forum hits and two times in resource hits in course 2 compare to course 1).

- Students are motivated to learn under teacher authoritative learning environment (more access hits in course 2 than course 1).

- More than one third of the students are lacking in written communication skill (cluster 0 in course 1 and cluster 2 and 3 in course 2).

- Learning environment differentiation can change the students online access behaviour. (Fig. 6 and Fig. 7)

## VI. Conclusion

From this research with the quantitative analysis, it could be concluded that the majority of the student population are not self-motivated to do self-learning. The highest number of students still lack the interest and motivation on accessing the e-learning site in parallel to their college studies. The existing motivation can be enhanced by grading and certification. Continuous guidance and support will provide a significant support to increase the student SDL motivation. However, it will not be an easy task to achieve the maximum effectiveness from the self-regulatory teaching with this learning behaviour of the students. Students have to be educated and their perception has to change on self-learning. Further, there may be some other parallel training sessions needed to develop the students' communication skills. There may be several other reasons for the higher number of students having less interest on using on-line self-directed learning materials. Poor quality of e-learning materials, language and technology incompetence, technical problems, system problems, inadequate facilities, isolation and inadequate support might be some reasons preventing the students from being the most active self-learners [15]. Further research is required to reveal the exact reasons for this situation and adequate solutions might change the students' attitude towards on-line Self-Directed Learning.